# Realizzazione di un Doppio Quadrante Astronomico Didattico per la XIII Olimpiade Internazionale di Astronomia[1]

## *The Double Didactic Astronomical Quadrant for the XIII International Astronomical Olympiad*

Michele Maris[1,2],
Claudio Corte[1], Conrad Böehm[1], Giulia Iafrate[1],
Massimo Ramella[1]

[1]INAF – Osservatorio Astronomico di Trieste
Via G.B. Tiepolo, 11
I 34131 - Trieste, Italy

email: surname@oats.inaf.it

---

## Abstract

In questo contributo illustriamo il progetto di una versione semplificata di doppio quadrante astronomico, progettato per fini didattici e realizzato in occasione dello svolgimento della gara osservativa delle XIII Olimpiadi Internazionali di Astronomia (XIII International Astronomy Olympiad, XIII IAO), Trieste (I), 13 – 21ottobre 2008.

Here we present the development of a simplified version of double astronomical quadrant, designed for educational aims and realized on the occasion of the observational round of the XIII International Astronomy Olympiad, held in Trieste (Italy) October 13-21, 2008.

Bilingual Italian/English
english version at the end of the document

Version 1.0 - 2010 Feb 26

Doppio Quadrante Astronomico Didattico, Maris et al, 2009

## 1  Le XIII Olimpiadi Internazionali di Astronomia (XIII IAO)

Le Olimpiadi Internazionali dell'Astronomia (IAO) costituiscono il culmine delle Olimpiadi Nazionali dell'Astronomia, e delle Olimpiadi Regionali. Le Olimpiadi si tengono a cadenza annuale e sono rivolte a studenti e studentesse tra i 14 e I 15 anni (Juniores) e 16 e 17 anni (Seniors). Le XIII IAO si sono svolte a Trieste dal 13 al 21 Ottobre 2008 e hanno impegnato un centinaio di ragazzi e ragazze provenienti da 19 paesi. Da regolamento le prove delle IAO sono divise in tre categorie: teorica, pratica e osservativa (lo stesso schema viene seguito per le selezioni regionali e nazionali). Le prove delle prime due categorie sono rispettivamente di carattere teorico/descrittivo e di problem solving astronomico. La gara osservativa invece ha lo scopo di verificare il grado di familiarità dei concorrenti con la volta celeste, i sistemi di coordinate astronomiche, gli oggetti presenti e alcuni semplici metodi di misura astronomica .

La XIII edizione delle IAO è stata la prima in cui l'uso di uno strumento è stato proposto e approvato per la gara osservativa. La scelta dei tipi di strumento: un doppio quadrante astronomico e un semplice cerchio azimutale astronomico, e del tipo di gara, sono state dettata dalla volontà di proporre un'esercitazione di astronomia pratica che onorasse tanto la tradizione marittima quanto quella astronomica della città di Trieste.

Per maggiori informazioni sulle Olimpiadi Italiane e Internazionali di Astronomia visitare il i siti: http://www.olimpiadiastronomia.it e http://www.issp.ac.ru/iao.

Doppio Quadrante Astronomico Didattico, Maris et al, 2009

## 2 Requisiti di Progetto

I requisiti per la progettazione del doppio quadrante, dettati dalle esigenze della gara, sono:

1. semplicità d'uso,
2. leggerezza,
3. sicurezza,
4. usabilità con entrambe le mani,
5. facile riproducibilità
6. accuratezza dell'ordine del grado.

Inoltre si voleva ottenere un valido strumento didattico, proponibile al di fuori di un Osservatorio si è scelto di utilizzare materiali economici, di facile reperibilità e metodi di lavorazione alla portata di una scuola a indirizzo tecnico/professionale. Oltre alla costruzione, le esigenze di gara hanno richiesto una fase di caratterizzazione degli strumenti necessaria per assicurare che i 30 esemplari prodotti fossero equivalenti onde evitare discriminazioni tra i concorrenti.

## 3 Descrizione

Lo strumento, visibile in Fig. 1, è costituito da un corpo centrale realizzato tagliando un profilato rettangolare di alluminio, che supporta un goniometro commerciale diviso in 180 gradi sessaggesimali. Le linee di mira sono state ottenute fissando ai lati opposti del corpo due tubetti di alluminio montati in tappi rettangolari di plastica opportunamente forati (A, B). L'interno del corpo e' dipinto di nero. Una sottile vite di bronzo sostiene l'alidada costituita da una barra di plexiglass forata, al cui centro e' stata incisa una linea di fede successivamente colorata con vernice nera. All'estremità inferiore dell'alidata è fissato con una vite un piombo da pesca che fornisce una forza peso sufficiente a garantire che l'alidada non venga influenzata dall'attrito della vite. In fase di taratura il peso può essere aggiustato per garantire la verticalità dell'alidada. La scelta dell'alidada di plexiglass, molto più impegnativa nella realizzazione del classico filo a piombo, è stata imposta dalla necessità di minimizzare la sensibilità dello strumento al vento. Dal lato opposto dell'alidada e' fissata una maniglia cilindrica che permette all'utilizzatore di sostenere lo strumento con una mano.

La lettura avviene reggendo lo strumento con una mano, accostando l'occhio alla linea di mira in modo da puntare l'oggetto di cui si vuole determinare l'altezza e abbandonando l'alidada in modo che sia libera di oscillare. Quando questa è ferma si può utilizzare l'altra mano per fermare l'alidada contro il quadrante e effettuare la lettura della scala graduata. Essndo la scala in gradi bidirezionale e i due tubetti costituenti la linea di mira identici, lo strumento può essere utilizzato sia reggendo lo strumento con la mano sinistra (modalità AB) sia reggendolo con la destra (modalità BA).

E' stato costruito un anello di plexiglass (non mostrato in figura) che permette di montare lo strumento in configurazione altazimutale su di un comune cavalletto per macchina fotografica. In questo caso è sufficiente fare la puntata e leggere il risultato sull'alidada.

La calibrazione è stata effettuata in due fasi:

1. messa a punto
    1. lo strumento è stato messo a bolla in orizzontale,

   2. la posizione del pesetto è stata variata fino a far cadere esattamente a metà della scala graduata la linea di fede l'alidada
   3. come ulteriore verifica si è ruotato lo strumento a ±45 gradi (usando delle sagome di riferimento) e si è controllato che l'alidada segnasse 45 gradi.

2. caratterizzazione

   1. misura di altezze da una postazione fissa di una serie di marcatori posti ad altezza nota, utilizzando prima la lina di mira AB poi la linea di mira BA,
   2. misure di altezza della Luna.

La fase di caratterizzazione ha mostrato che tanto le differenze tra l'uso in modalità AB o BA quanto le differenze tra uno strumento e l'altro erano inferiori al mezzo grado.

## 4   La Gara

Il maltempo ha impedito lo svolgimento della prova osservativa, che ragioni organizzative non hanno permesso di rinviare. Per questo motivo abbiamo potuto solo verificare le impressioni d'uso dei partecipanti durante una sessione pomeridiana non competitiva. I ragazzi si sono dimostrati molto interessati all'uso di questo strumento (numerosi partecipanti hanno chiesto di acquistarne un esemplare), dimostrando di poter utilizzare il doppio quadrante senza particolari difficoltà. Di conseguenza riteniamo che, al di là dell'aspetto "sportivo", questo tipo di strumento possa essere usato per vari tipi di attività didattiche sul campo, atte a illustrare in modo pratico gli aspetti basilari dell'astronomia di posizione.

Doppio Quadrante Astronomico Didattico, Maris et al, 2009

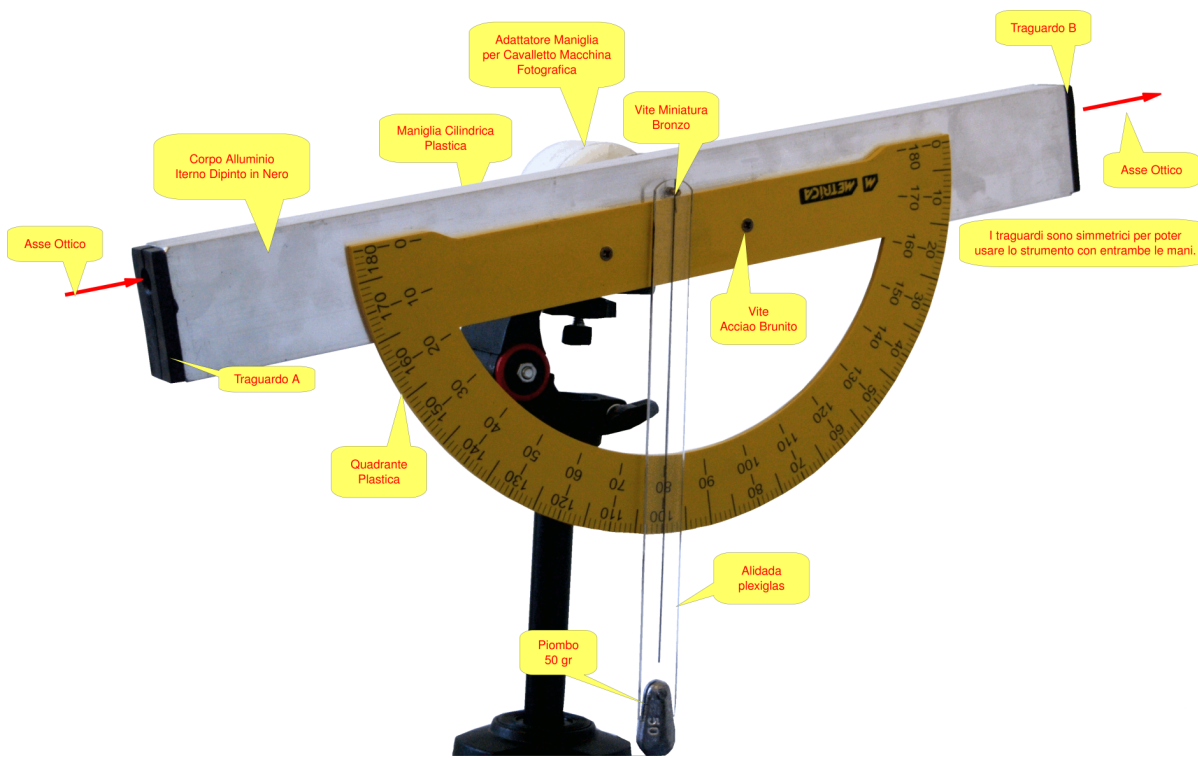

**Figura 1: Il Doppio Quadrante montato su un cavalletto per macchina fotografica.**

## 5  Altri Utilizzi

L'accoppiamento del doppio quadrante con un comune cavalletto per macchina fotografica dà la possibilatà di utilizzarlo come strumento dei passaggi per tracciare semplici carte astronomiche. Nell'ambito di un laboratorio di astronomia rivolto alle scuole secondarie di primo e secondo grado, questa attività permette di illustrare ai ragazzi i diversi sistemi di coordinate astronomiche (altazimutale ed equatoriale) con le relative trasformazioni, la relazione tra tempo solare e tempo siderale, i principi di tracciatura di una mappa celeste e il suo orientamento. Nel caso di una esercitazione con un sufficiente numero di studenti, si può inoltre illustrare come la combinazione di più misure indipendenti permetta di raggiungere una precisione migliore della precisione intrinseca allo strumento. In questo modo si può dimostrare sperimentalmente ai ragazzi come con una tecnologia relativamente semplice sia stato possibile, anche in epoche antiche, ottenere una buona accuratezza nelle misure trigonometriche e astrometriche, al contrario di quanto frequentemente affermato da alcuni media.

## 6  Ringraziamenti

Doppio Quadrante Astronomico Didattico, Maris et al, 2009

# English Translation

## 1  The XIII Internationa Astronomy Olimpiad (XIII IAO)

The International Astronomy Olmypiad (IAO) is an annual astronomy competition for high-school students, It is the last step of national and regional astronomy games involving 14-15 yr (junior) and 16-17 yr (senior) students. XIII IA0 occurred in Trieste in October 13-21, 2009, involving about a hundred boys and girls coming from 19 countries. According to rules the IAO rounds are divided into three categories: theoretical, practical and observational. It is worth to note that National and Regional selections follow the same scheme. The rounds of the first two categories are respectively of theoretical/desciptive and problem solving nature. The observational round aims instead to verify the participants knowledge of celestial sphere, astronomical coordinates systems, celestial onjects and some simple methods of astronomical measurements. During XIII IAO, for the first time, the use of an instrument has been proposed and approved for the observational round. The choice of instruments - a double astronomical quadrant and a simple azimuthal circle - and of the kind of match have originated from the wish to propose an astronomy exercise awarding the traditional heritage of the city of Trieste in astronomical navigation and science.

More information on the Italian and International Astronomy Olympiad are availble at http://www.olimpiadiastronomia.it and http://www.issp.ac.ru/iao/.

## 2 Design Requirements

The requirements for the double quadrant are drawn from the observational round rules. They are:

1. usage simplicity,
2. lightness,
3. safety,
4. possibility to be used with both left and right hand,
5. easy reproducibility
6. a precision of one degree.

Moreover, in order to have an effective educational tool, onw which could be proposed outside an astronomical observatory, we choose to use materials which are cheep and easy to found, while manufacturing techniques accessible to a technological/professional schools have been used. Beyond the production, the round rules required a characterization phase needed to ensure that the 30 pieces produced for the competition were equivalent, in order to avoid discriminations between the participants.

## 3 Description

The instrument, shown in Fig. 1, is based on a rectangular body, obtained by cutting a rectangular alluminium profile. It supports a commercial plastic protractor of large size and with two angular scales in sessaggesimal degrees.

The line of sight is defined by two short chuncks of alluminium pipe fixed at the opposite sides of the body by using two plastic caps with holes, marked in the figure A and B. The inside of the two chunks of pipe as the body are black painted.

A plexiglass bar acts as an alidada. The bar is fixed by a small bronze screw on the top, while at the bottom a small piece of lead is fixed, heavy enough to make any effect of friction on the alidada motion negligible. The lead could be rotated to set the exact zero point of the instrument. A black line is carved on the inner surface of the plexiglass bar and filled with black paint. The use of a more complex plexiglass alidada in place of a plumb bob, is motivated to reduce the sensitivity of thte instrument to the wind. On the opposite side a cilindrical handle is fixed at the body of the instrument to held the instrument with one hand.

The usage of the instrument is simple and intutive. One hand holds the handle. The lin of sight is pointed toward the object of which the height on the horizons has to be determined. The alidada is left free to oscillate, and it is stopped by the other hand to read the quadrant.

Being the whole instrument simmetrical along the line of sight the instrument could be held and used with the left hand (AB configuration) or with the right hand (BA configuration).

Eventually, we build a plastic ring to mount the instrument on a photographic tripod.

The instrument is calibrated in two steps:

La calibrazione è stata effettuata in due fasi:

1. calibration
    1. the instrument is oriented horizontally;

2. the lead is rotated to put at zero the line on the alidada;

3. a verification is done by using a reference plane tilted at 45 deg looking that the black line falls on the 45 deg division of the plastica scale.

2. characterization

   1. measurmente of a set of heights from a fixed position along a set of markers with known heigh, by using first the AB configuration and then the BA configuration;

   2. measures of heights of the moon.

Step 2 demonstrated that both differences between AB and BA configuration, as well as the differences among differente instruments where smaller than half degree.

## 4  The Competition

Bad weather prevent from performing the observing round and it have been impossible to pospone it. So that it have been just possibile to allow the partecipants to play with the instrument during a non competitive training session. Most of the participants found the instrument interesting and enjoy it without any particular problem. Consequently we conclude that this kind of instrument could be proposed for a number of astronomical didactic activities on field, in order to show in a practical manner basic aspects of positional astronomy and navigation.

## 5  Acknowledgements

The XIII IAO have been organized by INAF – Osservatorio Astronomico di Trieste under the auspicies of the International Committee for the Coordination of Euro-Asian

Astronomical Society (EAAS) and under the High Patronage of the President of Italian Repubblic, with the patronage and support of Italian Space Agency (ASI), of *Consorzio per la Fisica, della Regione Friuli-Venezia Giulia*, of *Provincia and Comune di Trieste*. Sponsors of the event have been *Assindustria* of Trieste and *Officine Meccaniche Vidali*.

The design, development, production and characterization of the quadrants have been carried out by the Mechanical Workshop and Laboratories of INAF-Osservatorio Astronomico di Trieste.